\documentclass[pop,fleqn]{w-art}
\usepackage{times}
\usepackage{w-thm}

\usepackage{mathrsfs}
\usepackage{amsmath,amssymb}
\usepackage{amsmath}
\usepackage{amscd}

\theoremstyle{plain}

\theoremstyle{definition}

\usepackage[]{graphicx}
\chardef\bslash=`\\ % p. 424, TeXbook

\newcommand{\bea}{\begin{eqnarray}}
\newcommand{\eea}{\end{eqnarray}}

\newcommand{\pint}{\makebox[0pt][l]{\hspace{2.4pt}$-$}\int}
\renewcommand{\v}{{\varphi}}
\renewcommand{\o}{{\omega}}
\newcommand{\atopfrac}[2]{\genfrac{}{}{0pt}{}{#1}{#2}}
\newcommand{\sfrac}[2]{{\textstyle\frac{#1}{#2}}}
\newcommand{\half}{\sfrac{1}{2}}

\newcommand{\Q}{\mathbf{Q}}
\newcommand{\E}{\mathbf{E}}
\newcommand{\q}{\mathbf{q}}
\newcommand{\Lop}{\mathscr{L}}
\newcommand{\contourgauge}{\mathbf{C}}

\newcommand{\alg}[1]{\mathfrak{#1}}

\newcommand{\su}{\alg{su}}

\def\pa{\partial}
\def\s{\sigma}
\def\l{\lambda}

\def\ads{{\rm AdS}_5\times {\rm S}^5}

\begin{document}
%%    The information for the title page will be placed between
%%    \begin{document} and \maketitle. The order of most entries
%%    is determined by the class file and can not be changed by
%%    rearranging them. The maketitle command follows after the
%%    abstract.
%%
%%    Most of the following commands will be completed by the publisher.
%%
%%    The copyrightyear is defined in the .clo file as the first argument
%%    of the copyrightinfo command. If the copyrightyear differs from that
%%    value it might be adjusted by the following definition:
%%
%% \renewcommand{\copyrightyear}{2003}% uncomment to change the copyrightyear.
%%
%\DOIsuffix{theDOIsuffix}
%%
%% issueinfo for header and copyright line
%\Volume{51}
%\Issue{1}
%\Month{01}
%\Year{2003}
%%
%%    First and last pagenumber of the article. If the option
%%    'autolastpage' is set (default) the second argument may be left empty.
\pagespan{1}{}
%%
%%    Dates will be filled in by the publisher. The 'reviseddate' and
%%    'dateposted' (Published online) entry may be left empty.
%\Receiveddate{~~~~~~~~~~}
%\Reviseddate{~~~~~~~~~~~}
%\Accepteddate{~~~~~~~~~}
%\Dateposted{~~~~~~~~}
%%

%% \pretitle{Editor's Choice}

%% We have a short and a long form for the title. The short form
%% (optional argument) goes into the running head.

\title[Quantum Strings and Bethe Equations]{Quantum Strings and Bethe Equations}

%% Please do not enter footnotes or \inst{}-notes into the optional
%% argument of the author command. The optional argument will go into
%% the header.  If there is only one address the marker \inst{x} may be
%% omitted.

%% Information for the first author.
\author[G. Arutyunov]{G. Arutyunov\footnote{Corresponding
     author: e-mail: {\sf agleb@aei.mpg.de}, Phone: +49\,0331\,567\,7229\,
     Fax: +49\,0331\,567\,7297}\inst{1}} \address[\inst{1}]{Max-Planck-Institut f\"ur Gravitationsphysik,
Albert-Einstein-Institut, \\
Am M\"uhlenberg 1, D-14476 Potsdam, Germany}
%%
%%    Information for the second author
%\author[Sh. Second Author]{L. Second Author\footnote{Second author footnote.}\inst{1,2}}
%\address[\inst{2}]{Second address}
%%
%%    Information for the third author
%\author[Sh. Third Author]{L. Third Author\footnote{Third author footnote.}\inst{2}}
%%
%%    \dedicatory{This is a dedicatory.}
\begin{abstract}
I briefly review the recently proposed construction of the Bethe ansatz which 
diagonalizes the Hamiltonian for quantum strings
on $\ads$ at large tension and restricted to the large charge states from a closed $\su(2)$ subsector.

\end{abstract}
%% maketitle must follow the abstract.
\maketitle                   % Produces the title.

%% If there is not enough space inside the running head
%% for all authors including the title you may provide
%% the leftmark in one of the following three forms:

%% \renewcommand{\leftmark}
%% {First Author: A Short Title}

%% \renewcommand{\leftmark}
%% {First Author and Second Author: A Short Title}

%% \renewcommand{\leftmark}
%% {First Author et al.: A Short Title}

%% \tableofcontents  % Produces the table of contents.

%%%%%%%%%%%%%%%%%%%%%%%%%%%%%%%%%%%%%%%%%%%%%%%%%%%%%%%%%%%%%%%%%%%%%%%%%%%%%
%%%%%%%%%%%%%%%%%%%%%%%%%%%%%%%%%%%%%%%%%%%%%%%%%%%%%%%%%%%%%%%%%%%%%%%%%%%%%%%
%%%%%%%%%%%%%%%%%%%%%%%%%%%%%%%%%%%%%%%%%%%%%%%%%%%%%%%%%%%%%%%%%%%%%%%%%%%%%%%%%

\section{Introduction}
Recently there has been a lot of effort to shed more light on the AdS/CFT duality 
conjecture by using the idea of exact integrability. 
On the gauge theory 
side this includes elucidation of integrable properties of the dilatation operator 
of the planar ${\cal N}=4$ SYM at the leading \cite{MZ,BS} and higher orders \cite{BKS} of perturbation theory
(see also \cite{QCD} for earlier account of integrable structures in QCD).  
On the string side recent developments are related to the study  \cite{FrT} of  
spinning strings in $\ads$ which provides new interesting information 
beyond the plane-wave limit \cite{BMN}. It appears that underlying integrability
of the string sigma model is indispensable 
for constructing explicit string solutions \cite{AFRT}.
Another important aspect concerns the 
near plane-wave quantization of strings. Here the main problem is to 
determine corrections to energies of the plane-wave states arising in the large curvature 
expansion \cite{PR,Callan}.

A relation between gauge and string theories can be probed by comparing
the scaling dimensions $\mathbf{\Delta}(\lambda)$ of the gauge theory operators ($\lambda$ is the 't Hooft coupling)
with energies $\mathbf{E}(\lambda)$ of the corresponding classical/quantum string configurations.
Since the planar dilatation operator admits interpretation as the Hamiltonian of an integrable (long-range)
spin chain the powerful method of the algebraic Bethe ansatz can be applied to compute $\mathbf{\Delta}(\lambda)$.
%To find $\mathbf{\Delta}(\lambda)$ The interpretation of the planar 
%dilatation operator as the Hamiltonian of an integrable (long-range)
%spin chain is crucial as it allows one to apply the powerful method of the algebraic Bethe ansatz.
Recently an all-loop asymptotic Bethe ansatz for the 
dilatation operator acting in the closed $\su(2)$ subsector was proposed \cite{BDS}. It provides a 
natural higher-loop generalization of the previously found the one-loop \cite{MZ}, two-loop and three-loop
Bethe ans\"atze \cite{SS} compatible with the {\it assumption} of the so-called BMN scaling.   

Quite generally, integrability implies the existence 
of a family of local commuting integrals of motion (charges) containing the Hamiltonian \cite{FT}.
Therefore, comparison of scaling dimensions
with string energies can be naturally extended to the whole towers of higher hidden
gauge/string charges \cite{AS}. The simplest string solutions correspond to rigid strings 
and they can be described in terms of the finite-dimensional integrable systems of the Neumann type \cite{AFRT}. 
More general (finite-gap) solutions are encoded in the integral equations of the Bethe type 
which are referred as the classical string Bethe equations (CSBE) \cite{KMMZ}. 

As is known,
the scaling dimensions of operators from the $\su(2)$ subsector  
agree with energies of rigid strings up to two-loop level \cite{BFMSTZ,AFRT,SS} but start to disagree
starting at three loops \cite{SS}. The same applies to the eigenvalues of all higher commuting charges 
\cite{AS}. 
Study of the curvature corrections to the plane-wave limit also reveals the similar pattern \cite{Callan}.
One hope for curing this disagreement is to take into account the so-called wrapping interactions \cite{BDS}.
By now, the match/mismatch of the gauge and string integrable structures can be shown 
in three different ways. The first one consists in computing the infinite tower of string charges 
by using the B\"acklund transform on rigid strings and further comparing it with that of the gauge theory \cite{AS} (see also \cite{E}).
The second is based on matching the CSBE with the gauge theory Bethe ans\"atze \cite{KMMZ}.
Finally, the third method uses the so-called ferromagnetic sigma model \cite{K}.        
A lot of important work on matching the particular gauge/string solutions as well as on 
understanding the integrable properties of gauge and string theories has been done. 
Unfortunately, we are not able to discuss it here 
and refer the reader to the original literature.  

The CSBE describing the finite-gap solutions of the string sigma model is an equation 
of the integral type \cite{KMMZ}. On the other hand, the gauge theory asymptotic Bethe ansatz (ABA) 
is a set of discrete (fundamental) equations. Assuming the validity of the AdS/CFT correspondence one should
expect existence of a Bethe ansatz for quantum strings which would
serve as a discretization of the integral (continuous) Bethe
equations for classical strings and, from the gauge theory
perspective, include terms responsible for wrapping interactions.
Recently a certain discretization of the CSBE has been proposed \cite{AFS}.
Here we briefly review the essential points of this construction, 
to which we refer as the quantum string Bethe equation (QSBE).
Before we proceed with technical issues let us summarize the properties 
of the proposed QSBE:
\begin{enumerate}
\item In the thermodynamic limit it describes the {\it classical} spinning strings. 
\item For the finite number of excitations (impurities), $M=2,3$, it reproduces the near plane-wave 
correction
to energies of the {\it quantum string} found 
in \cite{Callan} and gives a new prediction 
for any finite $M$.
\item At strong coupling it reproduces the $\sqrt[4]{\l}$
behavior of string energies/anomalous dimensions.
\item It agrees with gauge theory asymptotic Bethe ansatz up to two loops. 
\item It admits interpretation in terms of a long-range spin chain at weak coupling \cite{BC}.
\end{enumerate}
The general multi-impurity spectrum (in the $\su(2)$ subsector) predicted in \cite{AFS}
has been recently reproduced from the quantized string theory in the near plane-wave background \cite{McL}!
It would be important to extend the present construction to other closed subsectors 
(for certain extensions of the CSBE see \cite{KZ})and ultimately 
to the whole theory (see \cite{MS}). Perhaps, the formidable problem of {\it deriving} the
QSBE could be approached along the lines of \cite{AF}. In \cite{AF} we have obtained  
the classical bosonic Hamiltonian for strings on  $\ads$ in the uniform gauge,
showed its integrability and exhibited the corresponding spectral properties.
Finally, we note that it would be also interesting to understand the implications of the 
gauge theory ABA and the QSBE for the Operator Product Expansion in ${\cal N}=4$ SYM \cite{Arutyunov:2001mh}.

%%%%%%%%%%%%%%%%%%%%%%%%%%%%%%%%%%%%%%%%%%%%%%%%%%%%%%

\section{Classical String Bethe Equation}
A starting point in our construction of the
QSBE is the integral Bethe equation
which describes the finite-gap solutions of the classical string
sigma model. Therefore, it is useful here to recall its origin \cite{KMMZ}.

Consider a classical string moving in $\mathbb{R}\times {\rm
S}^3$. Here $\mathbb{R}$ stands for the global time direction in
${\rm AdS}_5$ and ${\rm S}^3$ is a three-sphere inside ${\rm
S}^5$. The non-vanishing embedding coordinates $X_1,\ldots, X_4$
are combined in a SU(2) matrix
\[
g=\left(
\begin{array}{cc}
X_1+iX_2 & X_3+i X_4 \\
-X_3+iX_4 & X_1-iX_2
\end{array}
\right) \, .
\]
This matrix is used to define an $\su(2)$-current
\[
A_{\tau}=g^{\dagger}\pa_{\tau}g, \qquad \qquad
A_{\sigma}=g^{\dagger}\pa_{\sigma}g \, ,
\]
because $X_i^2=1$. The Virasoro constraints read as   \[ {\rm
Tr}(A_{\tau}^2+A_{\sigma}^2)=-\frac{2\E^2}{\l}\, , \qquad \qquad
{\rm Tr}(A_{\tau}A_{\sigma})=0\, ,
\]
where $\E$ is the space-time energy of the string.

Introduce the $x$-dependent matrices ${\Lop}_{\s}$ and
${\Lop}_{\tau}$ (the Lax connection): \bea \nonumber
{\Lop}_{\s}=\frac{z}{1-z^2}A_{\tau}+\frac{1}{1-z^2}A_{\sigma} \,
,
%\\
%\nonumber
\qquad
{\Lop}_{\tau}
=
-\frac{1}{1-z^2}A_{\tau}-\frac{z}{1-z^2}A_{\sigma}\,
. \eea Here $z\in \mathbb{C}$ is the {\it spectral parameter}. One
can show that the equations of motion of the sigma model: \bea
\label{eom} \pa_{\tau}A_{\tau}-\pa_{\s}A_{\s}=0 \eea are
equivalent to the condition of zero curvature \cite{FT}
\[
[\mathscr{D}_{\tau},\mathscr{D}_{\sigma}]=0
\]
for covariant derivatives
$\mathscr{D}_{\tau}=\pa_{\tau}-{\Lop}_{\tau}$ and
$\mathscr{D}_{\sigma}=\pa_{\sigma}-{\Lop}_{\s}$.

The Lax connection can be used to construct an infinite tower of
integrals of motion for the evolution equations (\ref{eom}).
Define the {\it monodromy matrix} as the path-ordered exponential
of $\Lop_{\s}$:
\[
\mathscr{T}(z)=\mathscr{P}\exp \int_{0}^{2\pi} {\Lop}_{\s}
(z)\rm{d}\s \, .
\]
Using the condition of zero curvature one can easily show that
the spectral invariants of the monodromy matrix are conserved
under the time evolution. Let us parametrize the eigenvalues of
$\mathscr{T}(z)$ in the following way
\[
\mathscr{T}(z)=\left(
\begin{array}{cc}
e^{ip(z)} & \\
& e^{-ip(z)}
\end{array}\right)\, .
%\arccos \Big(\frac{1}{2}\rm{Tr}~T(\varphi) \Big)
\]
The function $p(z)$ is usually referred as {\it quasi-momentum}.
The basic idea of the finite-gap integration is to reconstruct
the quasi-momentum and, more generally, solutions of the evolution
equations by using the analyticity properties of $p(z)$. The
quasi-momentum has at least the same two first order poles at
$z=\pm 1$ as the Lax connection itself. The residues are $\pm
\frac{\pi \E}{\sqrt{\l}}$ (the dependence on $\E$ occurs due to the Virasoro constraints); they can be easily determined (up to
the sign ambiguity!) by diagonalizing the pole part of $\Lop_{\s}$.
We further assume that $p(z)$ has no other poles. This allows one
to define a resolvent
\[
\mathbf{G}(z)\equiv
p(z)+\frac{\E}{\sqrt{\l}}\frac{\pi}{z-1}+\frac{\E}{\sqrt{\l}}\frac{\pi}{z+1}\, ,
\]
which is a pole-free analytic function on the complex $z$-plane
with a (finite) number of cuts. One can show that the asymptotics
of $p(z)$ around $z=0$ and $z=\infty$ are related to the Noether
charges of the global symmetry group ${\rm SU(2)}_{\rm L}\times
{\rm SU(2)}_{\rm R}$ which, in their turn, can be expressed via
the Dynkin labels of the corresponding representations. Below we
just state the asymptotic properties of $\mathbf{G}(z)$ referring
the reader to the original work \cite{KMMZ} for further details:
\bea
&&\mathbf{G}(z)\rightarrow\frac{2\pi(\E-L+2M)}{z\sqrt{\l}}+\ldots
, \hskip 1.5cm {\rm when} \hskip 0.5cm  x\to  \infty \, ,\nonumber\\
&&\mathbf{G}(z)\rightarrow2\pi m+z\frac{2\pi
(\E-L)}{\sqrt{\l}}+\ldots , \hskip 1.5cm {\rm when }\hskip 0.5cm
x\to  0\, . \nonumber\eea Here $m\in \mathbb{Z}$ is a winding
number.

The analytic function $\mathbf{G}(z)$ can be expressed via the
spectral density $\rho_{\mbox{\scriptsize{s}}}(z)$ as
\[
\mathbf{G}(z)=\int_\contourgauge \frac{{\rm
d}z'\rho_{\mbox{\scriptsize{s}}}(z')}{z-z'}\, .
\]
The spectral density is supported an a finite number of cuts
$\contourgauge$.

Finally, unimodularity of the monodromy matrix implies that
on every cut one has
\[
\nonumber p(z+i0)+p(z-i0)=2\pi n \, , \qquad n\in \mathbb{Z} \,
.\] Since
\[
\mathbf{G}(z+i0)+\mathbf{G}(z-i0)= \int_\contourgauge\Big(
\frac{\rho_{\mbox{\scriptsize{s}}}(z)}{z-z'+i0}+
\frac{\rho_{\mbox{\scriptsize{s}}}(z)}{z-z'-i0}\Big)=
2\pint_\contourgauge
\frac{\rho_{\mbox{\scriptsize{s}}}(z')}{z-z'} \, .
\]
one gets (for every cut) the integral equation of the Bethe type
\[
\nonumber \pint_\contourgauge
\frac{\rho_{\mbox{\scriptsize{s}}}(z')}{z-z'}=\frac{\E}{L}\frac{\pi
z}{z^2-1}+\pi n\, , \] which is the CSBE \cite{KMMZ}.

The final step consists in  changing the spectral parameter $z\to
z(\v)$ according to the rule  \cite{BDS} \bea z\to
z(\v)=\frac{\v+\sqrt{\v^2-4\omega^2}}{2\omega}\, , \qquad
\omega^2\equiv\frac{\lambda}{16\pi^2 L^2}. \eea This change of
variables leads to the standard (field-theoretic type)
normalization of the density
\[ \nonumber \int_{\contourgauge} d\v\
\rho_{\mbox{\scriptsize{s}}}(\v ) =\frac{M}{L}=\alpha \, , \]
while the CSBE acquires rather
complicated form \cite{BDS}
\bea \label{CSBE}
&&\pint_{\contourgauge}
\frac{d\varphi'\rho_{\mbox{\scriptsize{s}}}(\v')}{\varphi-\varphi'}
=\frac{1}{2\sqrt{\varphi^2-4\o^2}}+\pi
n+\\
\nonumber && ~~~~+\o^2\int_\contourgauge
\frac{d\varphi'\rho_{\mbox{\scriptsize{s}}}(\v')}{\sqrt{\varphi^2-4\o^2}
\sqrt{\varphi'^2-4\o^2}}
\frac{\v-\sqrt{\varphi^2-4\o^2}-\v'+\sqrt{\varphi'^2-4\o^2}}
{(\v+\sqrt{\varphi^2-4\o^2})(\v'+\sqrt{\varphi'^2-4\o^2})-4\o^2}\,
.
 \eea

\section{Quantum String Bethe Equation}
Let us assume that there exists a set of fundamental (discrete)
equations of the Bethe type which in the thermodynamic (continuum)
limit leads to the integral equation (\ref{CSBE}). In general, the
procedure of discretization is by no means unique. We show,
however, that there exists a distinguished set of fundamental
equations, which agrees with all our current knowledge about {\it
quantum} strings on ${\rm AdS}_5\times {\rm S}^5$.

A crucial observation consists in rewriting eq.(\ref{CSBE}) in
the following form \cite{AFS}
 \[
\pint_{\contourgauge}\frac{d\varphi'\rho_{\mbox{\scriptsize{s}}}(\v')}{\varphi-\varphi'}
=\pi n+\frac{1}{2}\q_1(\v)+\sum_{r=0}^\infty \o^{2r+4}
\q_{[r+3}(\v)\Q_{r+2]} \, .\] Here the bracket $[.,.]$ stands to
denote antisymmetrization of indices and we use the notation
\[\Q_r =
\int_\contourgauge\ d\v\ \rho_{\mbox{\scriptsize{s}}} (\v )\
\q_r(\v)\, , \qquad \q_r (\v) =
\frac{1}{\sqrt{\varphi^2-4\o^2}}\frac{1}
{\left(\frac{1}{2}\v+\frac{1}{2}\sqrt{\varphi^2-4\o^2}\right)^{r-1}}\,
.
\]
Now it is time to realize that the charges $\q_r(\v)$ are
precisely those which arise in the thermodynamic limit
($M,L\to\infty$, $M/L=\alpha$ fixed) from the local excitation
charges of the gauge theory ABA \cite{BDS}.
In the momentum basis these charges are \cite{BDS} \bea
\label{lc} \q_r(p)=\frac{2\sin(\frac{r-1}{2}p)}{r-1}
\left(\frac{\sqrt{1+8g^2\sin^2(\frac{1}{2}p)}\, -\, 1
}{2g^2\sin(\frac{1}{2}p)} \right)^{r-1}\, , \qquad g^2 =
{\frac{\lambda}{8\pi^2}}\, .\eea In particular, \bea \q_1(p)&=&p~~
\Leftarrow~~\mbox{Momentum}\, ,
\nonumber\\
\q_2(p)&=&\frac{1}{g^2}\Big(\sqrt{1+8g^2\sin^2(\half p)}-1\Big)
\Leftarrow \mbox{Energy}\, . \nonumber\eea To rewrite this
charges in terms of the variable $\v$ one has to invert the phase
function of ABA \cite{BDS}
\[
\varphi(p)=\half\cot(\half p )\sqrt{1+8g^2\sin^2(\half p)} \, .
\]
Such a remarkable connection between classical string and quantum
gauge theory emerging in the thermodynamic limit can be further
used to infer a possible fundamental form of the string Bethe
equations.

The Bethe equations we propose to describe the leading quantum
effects for strings in the $\su(2)$ sector are of the form \cite{AFS} 
\bea
\label{QSBE} 
\exp(iLp_k) = \prod_{\textstyle\atopfrac{j=1}{j\neq
k}}^M {\rm S}(p_k,p_j)\, , \qquad\sum_{k=1}^Mp_k=0 \, . 
\eea 
Here ${\rm S}(p_k,p_j)$ is the S-matrix for pairwise
scattering of local excitations with momenta $p_k$: \bea
\nonumber {\rm S}(p_k,p_j)=\underbrace{
 \frac{\varphi(p_k)-\varphi(p_j)+i}{\varphi(p_k)-\varphi(p_j)-i}}_{\rm S-matrix~~of~~ABA}
 \exp\Big(2i\sum_{r=0}^{\infty}\Big({\frac{g^2}{2}}\Big)^{r+2}
~\q_{r+2}(p_{[k})~\q_{r+3}(p_{j]})\Big) \, . \nonumber \eea The
string S-matrix is a product of two terms. The first term involves
the phase function $\v(p)$ and it is identical to the S-matrix of
the gauge theory ABA constructed in
\cite{BDS}. The second exponential term is built upon
the local excitation charges eq.(\ref{lc}) of gauge theory. It is
precisely this term which makes a difference between gauge
and string theory! It starts to contribute to the S-matrix
at order $\lambda^2$ which, from the point of view of gauge
theory, means the three-loop level of perturbation theory.

One should view eqs.(\ref{QSBE}) as the system of equations to
determine $M$ individual momenta $p_k$. As momenta are
found they can be further used to compute the string energy
\bea
\nonumber
 \mathbf{E}(g)=L+g^2 \sum_{k=1}^M \q_2(p_k)\, ,
\eea as well as the higher charges, $\Q_r=\sum_{k=1}^M\q_r(p_k)$.
Finally, we note that taking the logarithms of eqs.(\ref{QSBE})
and passing through the standard rooting  of performing the
thermodynamic limit we indeed recover the integral equation
(\ref{CSBE}).

The properties of the QSBE have been already listed in the Introduction and their derivation can be found in \cite{AFS}.
Here we note that the QSBE can be easily solved in the $1/J$
expansion, where $J=L-M$.  Up to the order $1/J$ 
the string energy is
\bea\nonumber
\E=J+\sum_{k=1}^M \sqrt{1+\lambda'\,n_k^2}-\frac{\lambda'}{J}
\sum_{\textstyle\atopfrac{k,j=1}{j\neq k}}^M \frac{n_k}{n_k-n_j}
\left(
n_j^2+n_k^2\sqrt{\frac{1+\lambda'\,n_j^2}{1+\lambda'\,n_k^2}}
\right)\, ,\eea
where $n_k\neq n_j$ are the mode numbers of local excitations.
This formula beautifully agrees with the near-plane wave spectrum of the quantized string theory \cite{McL}. 
The same remains true also for the case of the confluent mode numbers, see \cite{AFS} for details.

\section*{Acknowledgments}
I would like to thank  Sergey Frolov and Matthias Staudacher 
for enjoyable collaboration on the subject presented here.
I'm grateful to Niklas Beisert and Arkady Tseytlin
for many interesting discussions.
I also thank the organizers and the participants of 
the RTN-EXT Workshop in Kolymbari for the stimulating atmosphere
and interesting discussions.
The work was supported in part
by the European Commission RTN programme HPRN-CT-2000-00131 and by
RFBI grant N02-01-00695.

\end{document}